\documentclass[aps,pra,twocolumn,amsmath,amssymb,amsfonts,superscriptaddress,floatfix,nofootinbib]{revtex4-1}
\usepackage{epsfig} \usepackage{graphicx}
\usepackage[pdftex,bookmarks]{hyperref}
\usepackage[pdftex]{color}
\usepackage[table]{xcolor}  
\usepackage{float} 

\newcommand{\bgar}{\begin{eqnarray}}
\newcommand{\enar}{\end{eqnarray}} 
 
 \newcommand{\be}{\begin{equation}}
\newcommand{\ee}{\end{equation}}  
 \def\mincirc{\lower
  3pt\hbox{$\buildrel<\over{\hbox{$\mathchar"218$}}$}}
\newcommand{\eotvos}{E$\ddot{\rm o}$tv$\ddot{\rm o}$s}
\newcommand{\eotwash}{E$\ddot{\rm o}$t-Wash}
\newcommand{\ms}{MICROSCOPE}

\begin{document}

\title{\bf Testing the Equivalence Principle in space after the MICROSCOPE mission\\}

 \date{\today}

\author{Anna M. Nobili}
\affiliation{
Dept. of Physics ``E. Fermi'', University of Pisa, Largo B. Pontecorvo 3,
56127 Pisa, Italy}
\affiliation{INFN-Istituto Nazionale di Fisica Nucleare, Sezione di Pisa, Largo B. Pontecorvo 3,  56127 Pisa, Italy}

\author{Alberto  Anselmi}
\affiliation{Thales Alenia Space Italia, Strada Antica di Collegno 253, 10146 Torino, Italy}

\begin{abstract}
%
Tests of the Weak Equivalence Principle (WEP) can reveal a new, composition dependent, force of nature or disprove many models of new physics. For the first time such a test is being successfully carried out in space by the MICROSCOPE satellite. 
Early results show no violation of the WEP sourced by the Earth for Pt and Ti test masses with random errors (after $8.26\,\rm d$ of integration time) of about $1$ part in $10^{14}$,  and systematic errors of the same magnitude. 
This result improves by about $10$ times over the best ground tests with rotating torsion balances despite $70$ times less sensitivity to differential accelerations, thanks to the much stronger driving signal in orbit. 
The measurement is limited by  thermal noise from internal damping in the gold wires used for electrical grounding, related to their fabrication and clamping. 
This noise  was shown to decrease when the spacecraft was set to rotate faster than planned.
The result will improve by the end of the mission, as thermal noise decreases with more data. Not so systematic errors.
We investigate major non-gravitational  effects and find that \ms's  ``zero-check'' sensor, with test masses both made of Pt, does not  allow their separation from the signal. The early test  reports an upper limit of systematic errors in the Pt-Ti sensor which are not detected in the Pt-Pt  one,  hence would not be distinguished from a violation. 
Once all the integration time available is used to reduce random noise there will be no time left to check systematics. 
MICROSCOPE demonstrates the huge potential of space for WEP tests of very high precision and indicates how to reach it. 
To realize the potential, a new experiment needs the spacecraft to be in rapid, stable rotation around the symmetry axis (by conservation of angular momentum), needs high quality state-of-the-art mechanical suspensions as in the most precise gravitational experiments on ground, and must  allow multiple checks to discriminate a violation signal from systematic errors.
The design of the ``Galileo Galilei'' (GG) experiment, aiming to test the WEP to $1$ part in $10^{17}$  unites all the needed features, indicating that a quantum leap  in space is possible provided the new experiment heeds the lessons of \ms.

\end{abstract}

\maketitle

\section{Introduction}
\label{Sec:Introduction}

The General theory of Relativity (GR) stands on the fundamental assumption that in a gravitational field all bodies fall with the same acceleration regardless of their mass and composition, a    ``fact of nature''  known as the Universality of Free Fall (UFF) or the Weak Equivalence Principle (WEP).  The WEP is at the crossroads of the open problems of fundamental physics: the relation of quantum fields and gravitation; the nature of dark matter and dark energy; the absolute character of the fundamental constants of physics. Tests of the WEP provide severe constraints to ``new physics'' attempting to cross the gap between GR and the Standard Model of particle physics, or make sense of dark matter and dark energy.

The dimensionless \eotvos\,\cite{Eotvos1890} parameter
\begin{equation}\label{Eq:EtaEotvos}
\eta_{_{E\ddot otv\ddot os}}=\frac{\Delta a}{a}      
\end{equation}
quantifies the level of violation. $\Delta a$  is the differential acceleration measured between two test masses of different composition as they fall in the field of a source body with average acceleration $a$ (the so called ``driving signal'') . 

A reliable measurement of $\eta_{_{E\ddot otv\ddot os}}\neq0$ would amount to the discovery of a new long-range composition dependent force of nature and make a revolution in physics;  the higher the precision of the test, the higher the chances to find new physics. Conversely, the more sensitive the test yielding $\eta_{_{E\ddot otv\ddot os}}=0$,  the greater the fine tuning required for many physical models and theories to survive.

The best experiments, carried out by the \eotwash\ group with slowly Rotating Torsion Balances (RTB),  have established that there is no violation to about $1$  part in $10^{13}$\,\cite{Adel2008,TBfocusIssue2012}. 
While some improvement is still possible, gaining orders of magnitude  requires moving the experiment to a laboratory in space (see\,\cite{BraginskyPaper2017} and references therein). An experiment to test the WEP in orbit, named STEP, has been studied    since the 1970s\,\cite{WordenSTEP1973,WordenSTEPthesis1976,Worden1978}; in the 1990s, following the interest raised by a re-analysis of the \eotvos\ experiment\,\cite{Fischbach1986}, ESA and NASA  have investigated the mission in considerable  detail with the goal of testing the equivalence principle to $10^{-17}$\,\cite{STEPESA1993}.

For  the first time an equivalence principle  test is  carried out with test masses in low Earth orbit, weakly suspended inside the \ms\ spacecraft aiming to reach $10^{-15}$\,\cite{MicroscopeFocusIssue2012}. 
With the signal at  a few $\rm mHz$, MICROSCOPE scientists report for a Pt-Ti composition dipole a null result relative to the Earth with a random noise  on the \eotvos\ parameter $\eta_{_{\oplus}}$ of about  $10^{-14}$ (after an integration time of $8.26\,\rm d$), and systematic errors at the same level\,\cite{MicroscopePRL171204}. 

In the field of the Earth \ms's early result is   a $10$-fold improvement over rotating torsion balances.
The improvement occurs despite a sensitivity to differential accelerations about $70$ times worse than RTB at similar  frequency. In favor of the space test  is the much larger driving signal from Earth  by almost $500$ times   at low  altitude as compared to RTB on ground ($\sim8\,\rm ms^{-2}$ versus $0.0169\,\rm ms^{-2}$ at most)\,\cite{BraginskyPaper2017}. 
Having RTB   superseded  mass dropping  tests by several orders of magnitude, despite a driving signal almost $600$ times weaker ($\lesssim0.0169\,\rm ms^{-2}$ versus $9.8\,\rm ms^{-2}$)   the very large factor yet to gain in low Earth orbit and the success of \ms\ strongly indicate that the next leaps in precision tests of the WEP shall occur  in space. 


There is no such gain  over RTB  if  data  are analyzed  taking the  Sun, or  the dark matter in our galaxy, as  source bodies   of a possible WEP violation. In this case the gravitational and inertial forces which are being compared  are the gravitational attraction from the source body (either the Sun or dark matter at the center of the galaxy) and the centrifugal force that keeps the test masses in orbit around them, and there is no larger driving signal in low Earth orbit.  The best null results  in the field of the Sun and of dark matter have been established by RTB at $\eta_{_{\odot}}$ and $\eta_{_{DM}}$ of a few times $10^{-13}$ and a few times $10^{-5}$ respectively (\cite{TBfocusIssue2012}, Table\,3). With a sensitivity to differential accelerations  worse than RTB, no improvement can come from \ms.
 
 
 Candidate dark matter particles are typically new particles, not included in the Standard Model of particle physics, which would generate a long-range composition-dependent scalar interaction, hence violate the WEP.  RTB tests rule out such a new composition dependent interaction between dark matter and ordinary matter to a few parts in $10^{5}$\,\cite{TBfocusIssue2012} stating that, to this level of precision, dark matter in our galaxy interacts with ordinary matter via the gravitational interaction only. Our current understanding of the cosmos is based on the assumption  that the {required} non luminous dark matter interacts with ordinary matter only gravitationally and there is no new long-range interaction. Although this assumption is very often taken for granted, we should be reminded that it is only an assumption and as such it should be tested by the most sensitive possible experiments.


%
In \ms\ each  test cylinder is actively  controlled by electrostatic forces (electrostatic suspensions act as a negative spring, hence un-controlled cylinders  would be unstable\,\cite{Onera1999}) and the readout is capacitive too. Electric grounding is ensured by connecting each cylinder to the cage with a  thin gold wire which should have only an ancillary role. Nonetheless, the measurement is  limited by thermal noise due to  losses in the wires related to their fabrication and clamping.   
%

Thermal noise due to internal damping in the  suspensions  decreases with the frequency as $1/\sqrt{\nu}$\,\cite{Saulson1990, PRLthermalnoise}. 
A signal of  WEP  violation   with the Earth as source  would be   DC on ground and at orbital frequency in space. A way to increase this frequency is by rotating the sensor relative to the  Earth, the faster the better.   


In \ms\ each test cylinder is sensitive in 1D, along its symmetry axis. Hence, rotation relative to  the Earth must occur around an axis perpendicular to the symmetry axis (\cite{MicroscopePRL171204}, Fig.\,1), which is the only stable axis against small perturbations. Therefore rotation was  planned to be slow, below $5\nu_{orb}$\,\cite{MicroscopeFocusIssue2012}, but it has been raised in order to reduce thermal noise since it turned out to be higher than expected. The result reported in\,\cite{MicroscopePRL171204} has been obtained at   $\nu_{spin}=17.5\nu_{orb}\simeq2.94\,\rm mHz$, and this is the current baseline.  This is the first demonstration of a high precision rotating experiment in space. Rotation of the whole spacecraft relative to inertial space, with no stator and no bearings, has very low noise and  is the key to the mission success.

By the end of the mission, with more data available, thermal noise, being random, will decrease and allow a WEP test  closer to the $10^{-15}$ original target of the mission. The upper limit of currently reported systematic errors is about $10$ times larger than that. They will not disappear or decrease  with more data, and  all systematics that eventually  were to emerge above random noise shall require very careful checking  in order to be separated with certainty from a possible violation signal.   

For this purpose \ms\  carries a second, ``zero-check''  equal composition accelerometer (named SUREF) with the test cylinders both made of Pt. Ideally, a violation signal should appear in the Pt-Ti sensor  and not in the Pt-Pt one, while systematic effects due to known physics should be detected by both sensors.
In this work we compare random noise and non-gravitational systematic effects in the two sensors showing that  the expected separation of the violation signal from systematic errors  does not occur. 


A thorough check of systematic errors cannot be avoided. It requires a sufficient number of measurements, all to  the  same precision, to be carried out in different physical conditions such that the different physical parameters involved  allow the signal to be distinguished from systematics on the basis of  their different signature, hence different dependence on these parameters.  

 \ms\ scientists  planned to  reach the mission  target $\eta_{_{\oplus}}=10^{-15}$  with an integration time corresponding to  $120$ orbits, so that in a $2$-yr mission duration there would be many such measurements, making it possible to check the result and possibly even improve it.
The published data show, however,  that in $120$ orbits the measurement is about a factor of $10$ short of the target. If all the remaining  integration time is used to reduce random noise, there will be none  left to check systematics.

The only way to resolve the ambiguity between systematic errors and a possible violation signal is by flying another experiment with higher precision and a shorter integration time. 
If  care is taken in flying an experiment somewhat more sensitive than ground  balances, the improvement  achievable in space can be  impressive.
%
%
Rapid rotation  of the spacecraft and consequent  up-conversion of the signal to high frequency, at which thermal noise is low and integration time is short, are  the main drivers of the ``Galileo Galilei'' (GG) space experiment  to test the equivalence principle to $10^{-17}$ without invoking cryogenics (\cite{PRLthermalnoise,GGfocusIssue2012,GGwebpage,GGssoPLA2003}, \cite{BraginskyPaper2017}).

The paper is organized as follows.

In Sec.\,\ref{Sec:SuccessLimitations}   we present the early \ms\ results, compare them with RTB, show how  faster rotation has reduced thermal noise, discuss the reported systematic errors  and their current limited understanding.

In  Sec.\,\ref{Sec:SUEPvsSUREF} we quantitatively compare the effects of thermal noise from internal damping and of major systematic errors in the two sensors,  to find that at a precision closer to the $10^{-15}$ mission target the Pt-Pt sensor will not allow a violation signal to be separated out.
With not enough  time left to confirm or rule out a violation, another experiment in space is needed; in Sec.\,\ref{Sec:LessonsFromMicroscope} we argue  that   \ms\ itself, through its success and limitations,  shows that a much more  precise test of the WEP in orbit is possible and points out the key changes to be made in order to achieve it. 

In Sec.\,\ref{Sec:Conclusions}  we draw the conclusions.

\section{MICROSCOPE first test of the equivalence principle in space: success, limitations  and open issues}
\label{Sec:SuccessLimitations}

While \ms\ is still in orbit, early results with an integration time of  $8.26\,\rm d$  yield, for Earth  as the source body and test masses made  of Pt  (with $10\%$ of Rh)  and Ti  (with $10\%$ of Al), a null result at $1\sigma$ level of\,\cite{MicroscopePRL171204}:
\begin{equation}\label{Eq:etaPtTi}
\eta_{_{\oplus}}(\rm{Pt,Ti})=[-1\pm9(\rm{stat})\pm9(\rm{syst})]\times10^{-15}   \ \ \ .                       
\end{equation}
It  has been obtained with the spacecraft in low Earth orbit  at frequency  $\nu_{orb}=0.16818\,\rm mHz$ ($P_{orb}=5946\,\rm s$) rotating at $\nu_{spin}=2.9432\,\rm mHz$ ($P_{spin}=339.8\,\rm s$), whereby  a WEP violation signal would occur at  $\nu_{_{EP}}=\nu_{spin}+\nu_{orb}=3.1113\,\rm mHz$ ($P_{_{EP}}=321.4\,\rm s$).

By comparison with the best tests of WEP achieved on ground by RTB\,\cite{Adel2008,TBfocusIssue2012} this is a $10$-fold improvement. 
The improvement  occurs with a sensitivity to differential accelerations, at the signal frequency, of: 
\begin{equation}\label{Eq:DifferentialAcceleration}
\Delta a_{Pt-Ti}\simeq9\times10^{-15} g(h)\simeq7.1\times10^{-14}\,\rm ms^{-2} \ ,                      
\end{equation}
where $g(h)\simeq7.9\,\rm ms^{-2}$ is the average gravitational acceleration from Earth at  $710\,\rm km$ altitude. Instead,  RTB are sensitive to   $\Delta a=10^{-15}\,\rm ms^{-2}$  at a  signal frequency of $0.84\,\rm mHz$ for both the Be-Ti and Be-Al composition dipoles tested (\cite{Adel2008,TBfocusIssue2012}, Table\,3). Despite $70$ times less sensitivity, \ms's early test     is 10 times better thanks to the larger driving signal from Earth in orbit versus RTB   on ground\,\cite{BraginskyPaper2017}.


In low Earth orbit, air drag and solar radiation pressure acting on the outer surface of the spacecraft give rise to an equal and opposite inertial acceleration on every test mass weakly suspended inside the spacecraft many orders of magnitude smaller than $1$-$g$, but also many orders of magnitude bigger than the target violation signal. The effect is common mode in principle, but in real experiments a differential residual remains much larger than the signal and at the same frequency.
For \ms\ the inertial acceleration resulting from  drag  is roughly  $7$ orders of magnitude bigger than the signal.

 At $1$-$g$ the torsion balance can reach a relative precision of $1$ part in $10^{13}$ in the differential effect of WEP violation having been built with tolerances of only $1$ part in $10^{5}$, thanks to its capability to reject common mode effects. In \ms\ the test cylinders are suspended individually (they do not form a balance) and their configuration is frozen after assembling. Any difference resulting from construction and mounting errors can only be mitigated (rejected) by in-flight calibrations of their individual responses. Matching by $8.5\times10^{-3}$  (rejection factor of about $118$) is reported in\,\cite{MicroscopePRL171204}. Such a low level of  rejection means that in order to achieve the result  (\ref{Eq:etaPtTi})  most of the drag acceleration has been successfully compensated by drag-free control of the spacecraft with propellant and thrusters. If combined with a comparable rejection  level by means of  an appropriate differential design of the sensor itself, it would  allow a much higher precision to be reached without stringent requirements on thruster noise\,\cite{STEPESA1996}.

 In \ms\ each test cylinder is weakly suspended along its symmetry axis by means of electrostatic forces, with $600\,\rm\mu m$ gap. Even after compensation of the large effect of drag, a restoring force is needed  in response to small residual forces  in order to prevent each  cylinder from hitting the cage. Since electrostatic suspensions are  unstable (they act as a  ``negative'' spring\,\cite{Onera1999}), a restoring force --including  the response to the violation signal, if any--  is provided by active electrostatic control.  
  
In this type of accelerometers developed in France at ONERA a loose, thin, conducting  wire made of gold ($7\,\rm\mu m$ width and $2.5\,\rm cm$ length in \ms\,\cite{MicroscopeBremen2017}) provides a physical connection between the test mass and the cage, primarily to avoid electric charging. 
 
 Concern about thermal noise from internal damping  in the wire at the low frequencies of interest has led  to extensive measurements of its quality factor $Q$ under realistic flight conditions.  An  \textit{ad hoc}, electrostatically suspended torsion pendulum was  built in order to avoid the  suspension wire and thus  achieve a very weak torsional constant\,\cite{WillemenotThesis,Willemenot2000a,Willemenot2000b}. The electrostatic pendulum is  in fact   $10$ times stiffer than the mechanical torsion pendulum of the  \eotwash\ group\,\cite{Adel2008}. It has measured the $Q$ of the gold wire  ($7.5\,\mu\rm m$ width and $1.7\,\rm cm$ length in this case) at frequencies ranging  from about $10^{-4}\,\rm Hz$ to several $10^{-2}\,\rm Hz$, showing the presence of large losses, with better $Q$  occurring at higher frequencies. The values measured range  from $Q=36$ slightly above $10^{-4}\,\rm Hz$ (the orbital frequency) to 59 at $10^{-3}\,\rm Hz$, while $Q$ values around $110$ are measured only at frequencies of $10^{-2}\,\rm Hz$ and a few times  $10^{-2}\,\rm Hz$(\cite{Willemenot2000a}, Fig.\,5).  
 
These  losses are  much higher than in the suspensions of small force ground experiments.  At a signal frequency slightly less than a $\rm mHz$, the \eotwash\ group reports a quality factor $100$ times better than measured by ONERA at $1\,\rm mHz$,   of about $6000$, with a $20\,\mu\rm m$ W wire suspending a $70$ gram balance at $1$-$g$\,\cite{Adel2008,TBfocusIssue2012}. Large losses also at low frequencies can be avoided, especially for suspensions to be used in weightlessness conditions, where even large masses need very low stiffness. 
Monolithic suspensions, manufactured from a single block (to avoid relaxation of bending energy) with enlarged ends (to ensure that clamping is located far from where the flexure undergoes deformation during motion), and with appropriate heat treatments, are commonplace in small force gravitational experiments and have low losses. Instead, a gold wire clamped with droplets of glue at its ends, where most dissipation occurs, is bound to yield large losses. Moreover,  losses will be unequal even if all wires  are taken from the same coil and assumed to be perfectly identical, because  the  procedure used for clamping is hardly repeatable. 

It has been known since   1990\,\cite{Saulson1990} that thermal noise from internal damping in the suspensions of the test masses in gravitational wave detectors  decreases with the frequency as $1/\sqrt{\nu}$.  This is how the Virgo/LIGO detectors around $100\,\rm Hz$ can be sensitive to   displacements of  the mechanically suspended mirrors as small as  about  $10^{-19}\,\rm m$\,\cite{LIGOfacts}. A signal of WEP violation  is at a much lower frequency: with  Earth as the source body, it  is DC on ground and at orbital frequency in space. For this reason rotation of the apparatus is used, as for  torsion balances,  to up-convert the signal to higher frequencies, the higher the better.

Limitations to the spin rate of RTB come from concerns about rotation noise (on ground it includes motor and bearings noise) and the attenuation of the signal strength at frequencies above the natural  oscillation mode (the system being in essence a forced oscillator\,\cite{PRLthermalnoise}). With a natural torsional  frequency $\nu_{tor} =\frac{1}{798}\,\rm Hz$, the highest  spin rate so far is $\nu_{spinRTB}=\frac{2}{3}\nu_{tor}\simeq0.84\,\rm mHz$\,\cite{TBfocusIssue2012}.

In \ms\  rotation occurs perpendicularly to the orbit plane and  to the symmetry axis (\cite{MicroscopePRL171204}, Fig.\,1), which however is the only stable axis  against small perturbations. For this reason a slow rotation mode was planned with $3\nu_{orb}<\nu_{spin}<5\nu_{orb}$\,\cite{MicroscopeFocusIssue2012}.
Once in orbit,  thermal noise turned out to be higher than expected; being dominated by internal damping, the cure was to reduce it by increasing  the rotation rate of the spacecraft. The  result (\ref{Eq:etaPtTi}) was obtained while spinning at $\nu_{spin}=17.5\nu_{ord}\simeq2.94\,\rm mHz$, which is $3.5$ times faster than the maximum spin rate planned before launch. The possibility --unique to space-- of rotating the whole spacecraft, with no stator and no  bearings, has proved to be much less noisy than  rotating experiments in ground laboratories. The  faster rotation rate  has been adopted as the new baseline, despite the higher consumption of propellant and consequent shorter duration of the mission.  

At a recent conference on ``Fundamental Physics in Space'' in Bremen \ms\ scientists have reported that when up-converting the signal frequency by increasing the rotation rate of the spacecraft, thermal noise in the Pt-Ti accelerometer improves more than expected according to the $1/\sqrt{\nu}$ dependence\,\cite{MicroscopeBremen2017}. The measurements show that while the signal frequency  increases by a factor $3.53$, hence thermal noise from internal damping is expected to decrease by $\sqrt{3.53}=1.88$,  it is instead found to decrease by $3.61$ times, with an unexplained (favorable) factor of $1.9$. 
The acceleration due to thermal noise from internal damping depends on the quality factor as $1/\sqrt{Q}$ (see Eq.\,(\ref{Eq:AccelerationThermalNoiseID}), Sec.\,\ref{Sec:SUEPvsSUREF}), and the quality factor too has been found to  depend on the frequency, usually being higher (lower losses) at higher frequencies. If so, the unexplained improvement by a factor $1.9$  might be due to an increase of the quality factor of the gold wires by  $1.9^{2}=3.6$ times for the same system in the same conditions, except for the fact that the frequency of the signal has increased (with faster rotation) by $3.53$ times.
 It is quite interesting that a similar  improvement has been observed with the ground demonstrator of the proposed GG experiment in space: with a frequency increase by $2.16$ times, the quality factor (for the same system, except for rotation), was found to increase by $2.24$ times (\cite{BraginskyPaper2017}, Sec.\,8).

The systematic errors shown  in (\ref{Eq:etaPtTi}) have been found to be mostly of thermal origin. 
The electronics unit and the base plate of the Pt-Ti sensor were subjected to artificially produced thermal variations at the signal frequency, and the resulting differential acceleration between the two  cylinders was measured, thus mapping the sensitivity of the instrument to thermal effects. Effects due to thermal variations of the base plate turned out to dominate over those of the electronics unit. But  they were  larger than expected by more than two orders of magnitude per degree of temperature variation, and the reasons for such behaviour are not known yet. At the same time, by measuring the actual thermal variations  (aside from those induced for the sensitivity test), it turned out that they were smaller than expected also by about two orders of magnitude\,\cite{MicroscopePRL171204}! Both findings call for a convincing physical explanation which may require more information, in particular on the residual pressure.  \ms\ scientists expect  that thermal stability can be even better  and therefore consider the systematic error reported in (2) as an upper bound.

A rotation rate faster than planned may be responsible for a thermal stability better than expected, because of a better averaging and also because  the signal frequency is farther away from the orbital frequency, at which most of the thermal stress obviously occurs.

\section{Thermal noise, systematic errors and the ``zero-check''  sensor}
\label{Sec:SUEPvsSUREF}

In addition to the composition dipole SUEP, whose inner (denser) cylinder is made of Pt  (with $10\%$ of Rh) and the outer one of Ti  (with $10\%$ of Al), \ms\ carries a second sensor, named SUREF, with  the same geometry (and the same $600\,\mu\rm m$ gaps) as SUEP but  cylinders made of the same material. The  inner cylinder is ``identical'' to the inner one in SUEP, being made of the same Pt-Rh  alloy (the two masses differ only by a few parts in $10^{4}$);  the outer cylinder has the same size and volume as the outer one in SUEP, but  it  is  made of Pt-Rh alloy like the inner one.  All densities are  uniform. 

The Pt-Pt SUREF sensor should  allow systematic errors to be distinguished  from a violation signal (``zero-check'' sensor):  a  violation signal --being composition dependent--  should appear in SUEP but not in SUREF, while systematic errors should appear in both sensors.

The two sensors are located $17.5\,\rm cm$ away from each other, and none of them is centered on the center of mass of the spacecraft. The drag-free control loop is closed  either  on one sensor or on the other,  with the task of partially compensating the common mode motion,  relative to the spacecraft,  of the two selected  cylinders due to the inertial acceleration (nominally the same for all test bodies) resulting from the effect of drag acting on the outer surface of the spacecraft. Depending on which sensor drives the drag-free control, science data are collected for that sensor only. Hence, SUEP and SUREF   never take data simultaneously, which weakens the role of SUREF as zero-check sensor and reduces the integration time available for SUEP to test the WEP. 

 \ms\ scientists expect, by the end of the mission, to accumulate sufficient data for the  random noise of SUEP shown in  (\ref{Eq:etaPtTi})  to reduce to a value close to the $10^{-15}$ target of the mission. Then, unless systematic errors will turn out to be much lower than the upper limit currently established, they will  emerge above noise and call for a full understanding of their physical origin, as they might contain a violation signal. 

The  Pt-Pt SUREF sensor should  detect only systematic effects due to known physics, not a violation signal, and thus solve the problem.
In reality, SUREF detection of systematic errors depends on its thermal noise (only those  larger than  thermal noise will be detected) and on its own sensitivity to systematics. If it turns out to be  less sensitive than SUEP to some systematics, it cannot rule them out completely as possible violation of the WEP.

The measurements are limited by  thermal noise from internal damping in the  gold wires. 
At the frequency of the violation signal $\nu_{_{EP}}$   the spectral density of the acceleration noise of each cylinder  (expressed in $\rm ms^{-2}/\sqrt{Hz}$ in SI units) reads\,\cite{Saulson1990}:
\begin{equation}\label{Eq:AccelerationThermalNoiseID}
\hat a_{_{thID}}=\frac{1}{\cal M}\sqrt{\frac{4K_{B}Tk_{w}}{Q_{w}\omega_{_{EP}}}}           
\end{equation}
where $K_{B}$ is the Boltzmann constant, $T$ is the equilibrium temperature, $\cal M$  is the mass of the test cylinder, $k_{w}$ and $Q_{w}$ are the stiffness and  quality factor of the gold wire connecting it to the cage, and $\omega_{_{EP}}=2\pi\nu_{_{EP}}$ is the frequency of the  signal.  Being random noise, and most probably uncorrelated  between  the inner and outer  cylinder in each sensor, the resulting differential acceleration noise competing with the signal is:
\begin{equation}\label{Eq:DifferentialAccelerationThermalNoiseID}
\hat {\Delta a}_{_{thID}}=\sqrt{\hat a_{_{thIDinner}}^{2} + \hat a_{_{thIDouter}}^{2}}   \ \ \ .       
\end{equation}
Assuming the same ambient temperature $T$  in the two sensors and the same  $k_{w}$ and  $Q_{w}$ for all gold wires (even at different frequencies), the ratio of the differential acceleration noise between   SUEP and SUREF  is:
\begin{equation}\label{Eq:RatioThermalNoiseIDSUEPvsSUREF}
\begin{split}
\frac{\hat {\Delta a}_{_{thID-SUEP}}}{\hat {\Delta a}_{_{thID-SUREF}}}=\\
=\sqrt{\frac{\omega_{_{EP-SUREF}}}{\omega_{_{EP-SUEP}}}}\cdot
\frac{\sqrt{\frac{1}{{\cal M}_{_{inner-SUEP}}^{2}}  +   \frac{1}{{\cal M}_{_{outer-SUEP}}^{2}}}  }{\sqrt{\frac{1}{{\cal M}_{_{inner-SUREF}}^{2}}  +   \frac{1}{{\cal M}_{_{outer-SUREF}}^{2}}}  }
\end{split}      
\end{equation}
showing that  it depends only on the different masses of the individual cylinders  and on the different  ratio between the frequencies of the signal, which in turn depends on the different rotation frequency during the respective measurements. The masses are measured  very precisely and  their contribution to the noise ratio (\ref{Eq:RatioThermalNoiseIDSUEPvsSUREF}) is $1.5989$. We must therefore expect a higher thermal noise in SUEP than in SUREF by about $1.6$, only because of the different values of the masses.  The recent measurements  for  SUEP and SUREF have been made at different rotation rates of the spacecraft, hence they refer to different  signal frequencies whose ratio is $3.1113\,\rm mHz/0.9250\,\rm mHz=3.3636$\,\cite{MicroscopePRL171204},  contributing to the  noise ratio (\ref{Eq:RatioThermalNoiseIDSUEPvsSUREF}) by $1/\sqrt{3.3636}=1/1.8340$. Overall we get: 
\begin{equation}\label{Eq:RatioThermalNoiseIDSUEPvsSUREF-numbers}
\frac{\hat {\Delta a}_{_{thID-SUEP}}}{\hat {\Delta a}_{_{thID-SUREF}}}=\frac{1.5989}{1.8340}\simeq0.87
\end{equation}
which means that, at the selected rotation frequencies and with the assumptions made we should expect, in the differential acceleration noise competing with the signal, a slightly lower  noise in SUEP than in SUREF.  To the contrary,  the  measured values reported in\,\cite{MicroscopePRL171204}  are $5.6\times10^{-11}\,\rm ms^{-2}/\sqrt{Hz}$ for SUEP and     $1.8\times10^{-11}\,\rm ms^{-2}/\sqrt{Hz}$  for SUREF,  showing  that SUEP is in fact $3.11$  times more noisy. This means that there is an unexplained factor of about $3.57$, SUEP being $3.57$ times  more noisy than  SUREF than expected. 


Since the temperature is well measured, only differences in the ratio $k_{w}/Q_{w}$ for the test cylinders  can be invoked, at least as long as random noise is due to internal damping as in (\ref{Eq:AccelerationThermalNoiseID}). Since all four wires have the same length and are cut from the same coil, while $Q_{w}$ depends mostly on the glue clamping at the two ends of each wire --which are unpredictable and hardly repeatable-- $Q_{w}$ is more likely to be responsible for the observed discrepancy. It appears in (\ref{Eq:AccelerationThermalNoiseID}) under the square root, hence the $Q_{w}$ which dominates thermal noise in SUEP should be almost $3.57^{2}\simeq13$ times smaller than the one which dominates in SUREF. 

Accelerometers similar, in their key features, to those of \ms, also built by ONERA, have successfully flown onboard the GOCE geodesy mission of the European Space Agency  and a noise level about two times larger than   expected has been reported in that case\,\cite{GOCEmicroscopeColloqium2013, TouboulmicroscopeColloqium2013,GOCEdataProcessing2011}. 

Concerning systematic effects at the frequency of the violation signal, and the respective sensitivities of SUEP and SUREF, we notice the following. The systematic errors which limit the \eotvos\ parameter (\ref{Eq:etaPtTi})  as measured with SUEP  are depicted in\,\cite{MicroscopePRL171204}, Fig.\,3 left plot (and  listed in Table III of the paper) at the level of about  $7\times10^{-14}\,\rm ms^{-2}$ as function of time with the number $N$ of orbits (120 in total). The same plot shows  the random acceleration noise, which instead decreases as $1/\sqrt{N}$ to meet, towards the end of the run, the horizontal line of systematic errors. The same Figure (right plot)  shows the (lower) random noise in SUREF, also decreasing as $1/\sqrt{N}$ over a total of $62$ orbits, to reach  slightly below $3\times10^{-14}\,\rm ms^{-2}$ at the end of the run.  
 Should there be systematic  effects as large as the upper limit reported in SUEP,  they would clearly appear above random noise in SUREF, but no such systematics are detected. They don't appear either in the spectral density of the acceleration noise of SUREF shown in\,\cite{MicroscopePRL171204}, Fig.\,2 right plot. The question as to why it is  so is obviously a very relevant one, because systematic errors should be detected by both SUEP and SUREF in order to be distinguished from a violation signal.
 
The systematic errors reported in SUEP and not detected in SUREF are non-gravitational. We therefore compare the sensitivity of SUEP and SUREF to non-gravitational perturbations. As shown in (\ref{Eq:EtaEotvos}), in WEP tests the physical observable is the differential acceleration of the test masses relative to the source body, hence, the relevant quantities are accelerations, not forces. The accelerations of a number of  non-gravitational perturbations are known to be proportional to the area-to-mass ratio of the affected body\,\cite{MNF}, the area being in this case that of the cross section of the test cylinder perpendicular to its sensitive/symmetry axis.
The differential acceleration  between the test cylinders in SUEP caused by such non-gravitational perturbation would be:
\begin{equation}\label{Eq:Deltaa_ngAsuMSUEP}
\begin{split}
\Delta a_{_{ng{\cal A/\cal M}-SUEP}}=\\
=a_{_{ng{\cal A/\cal M}outer-SUEP}}-a_{_{ng{\cal A/\cal M}inner-SUEP}}\propto\\
\propto{(\cal A/\cal M)}_{outerSUEP}-{(\cal A/\cal M)}_{innerSUEP}
\end{split}
\end{equation}
where ${(\cal A/\cal M)}$ is the area-to-mass ratio for the test cylinder referred to in the subscript; and similarly for SUREF.
If the physical parameters which determine the non-gravitational perturbation under consideration are the same in both sensors,   the ratio of the differential accelerations it gives rise to depends only on the ratios of ${(\cal A/\cal M)}$ for the test cylinders in the two sensors:
\begin{equation}\label{Eq:Deltaa_ngAsuMRatio}
\begin{split}
\frac{\Delta a_{_{ng{\cal A/\cal M}-SUEP}}}{\Delta a_{_{ng{\cal A/\cal M}-SUREF}}}=\\
=\frac{{(\cal A/\cal M)}_{innerSUEP}}{{(\cal A/\cal M)}_{innerSUREF}}\cdot\\
\cdot\frac{{{(\cal A/\cal M)}_{outerSUEP}}/{{(\cal A/\cal M)}_{innerSUEP}}-1}{{{(\cal A/\cal M)}_{outerSUREF}}/{{(\cal A/\cal M)}_{innerSUREF}}-1}\ \ \  \ \ .\\
\end{split}
\end{equation}
With data available  on the masses  and the geometry of the  test cylinders\,\cite{MicroscopePRL171204,TouboulMoriond2011,MicroscopeTestMassesPTB} we get for this ratio (in modulus) about $3.3$. Thus, non-gravitational perturbations whose accelerations  are proportional to the area-to-mass ratio of the test cylinders, give rise to differential accelerations $3.3$ times larger in SUEP than in SUREF, simply because of the way they have been designed, all other physical parameters being the same. 

Since the upper limit of systematic effects reported in SUEP at the frequency of the signal is of  $7\times10^{-14}\,\rm ms^{-2}$, we should expect systematics in SUREF to be a factor $3.3$ smaller, hence of about $2\times10^{-14}\,\rm ms^{-2}$ at most. Being below the thermal noise measured in SUREF (\cite{MicroscopePRL171204}, Fig.\,3 right plot) it is not surprising that they are not detected.

This fact questions the use of the zero-check Pt-Pt SUREF sensor to discriminate a violation signal from spurious effects. With a longer integration time and a lower thermal noise SUREF may allow these systematics to be detected. However, as long as their value in SUEP is several times  larger than in SUREF the open issue remains that they may contain a violation signal as well. 

Other non gravitational effects, such as electric charging, do not depend on the cross section but only on the inverse of the mass.  In this case the ratio of the differential accelerations between the test cylinders in the two sensors (all other physical parameters being the same) reads:
\begin{equation}\label{Eq:Deltaa_ngmRatio}
\begin{split}
\frac{\Delta a_{_{ng{\cal M}-SUEP}}}{\Delta a_{_{ng{\cal M}-SUREF}}}=\\
=\frac{({\cal M})_{innerSUREF}}{({\cal M})_{innerSUEP}}\cdot\\
\cdot\frac{({\cal M})_{innerSUEP}/({\cal M})_{outerSUEP}-1}{({\cal M})_{innerSUREF}/({\cal M})_{outerSUREF}-1}\\
\end{split}
\end{equation}
which yields $0.47$ (modulus), meaning that this kind of systematic errors would be about $2$ times larger in SUREF than in SUEP.  Hence, electric charge effects, if any, are below the level of thermal noise  reported in SUREF.  Were such a spurious effect, at some point, detected above thermal noise in SUEP, the ratio (\ref{Eq:Deltaa_ngmRatio})  in favor of SUREF would not prevent its separation from the signal unless the observed ratio between the two effects is as theoretically expected.

Another disturbance competing with the signal, that questions the use of SUREF as a zero-check sensor is  the radiometer effect, which is proportional to the residual pressure around the test cylinders and to the temperature gradient between the two ends of its sensitive/symmetry axis\,\cite{RadiometerPRD2001}. 

\ms\ scientists exclude radiometer as the origin of the systematic effect in SUEP because of the extremely good thermal stability and uniformity observed. A residual pressure of $10^{-5}\,\rm Pa$ was assumed before launch\,\cite{MicroscopeFocusIssue2012} but no value is given once in orbit. By comparison, LISA Pathfinder (LPF)  established an upper limit of $2.2\times10^{-5}\,\rm Pa$  at the start of the data take\,\cite{LISAPF2016}, and a few $10^{-6}\,\rm Pa$ at the end of the experiment, because of venting to outer space and more time available for degassing\,\cite{LISAPF2018}. There is no venting to outer space in \ms, and the getter pumps it relies upon after final assembling in order to take care of pressure increase due to outgassing surfaces, have a limited lifetime and are inefficient with noble gases. Thus, a reliable estimate of the residual pressure both in SUEP and in SUREF is needed, especially because different values may be expected due to the fact that the outer cylinder in SUEP is the only coated one of the four, and this may result in a different outgassing as compared to the outer cylinder in SUREF, despite the same geometry. 

The radiometer effect should be carefully investigated  because it is known to give no differential acceleration if the two cylinders have the same density, as in the case of SUREF\,\cite{radiometerMicroscope}.  We have checked this fact with the numbers available for the test  cylinders\,\cite{MicroscopePRL171204,TouboulMoriond2011}, and the result for the ratio of the radiometer acceleration on the outer and inner cylinder in SUREF (of length $L_{_{outer}}$ and $L_{_{inner}}$, as in SUEP) is almost exactly $1$: 
\begin{equation}\label{Eq:RatioRadiometerSUREF}
\begin{split}
\frac{a_{_{rad-outerSUREF}}}{a_{_{rad-innerSUREF}}}=
\frac{{{(\cal A/\cal M)}_{outerSUREF}}\cdot L_{_{outer}}}{{{(\cal A/\cal M)}_{innerSUREF}}\cdot L_{_{inner}}}=1.009
\end{split}
\end{equation}
while the same ratio  for SUEP is $4.562$.
Thus,  the radiometer effect gives rise to a non-zero differential acceleration in SUEP and no differential acceleration in SUREF,  just as one expects for the violation signal. As suggested in\,\cite{radiometerMicroscope}, a way out of this  impasse might have been to fabricate the Pt-Pt  cylinders in SUREF with a different average density, e.g. with some appropriate empty volume in one of them, e.g. the outer one. 

Although some additional information may be available (by investigating the accelerations as measured individually by each test cylinder) 
that could  help to mitigate the difficulties outlined here, it is likely that the main goal of the second equal composition sensor to provide a clear-cut, unquestionable, check of  the violation signal versus systematic errors will not be met.

If so, the need shall arise for carefully designed checks of systematic errors in the different composition sensor. This requires many measurements, all at the same sensitivity (possibly the target sensitivity) in different physical conditions, such that  systematic errors and violation signal can be distinguished on the basis of their respective signature and consequent different dependence on the physical parameters involved in these measurements, as it is done in ground tests of the WEP with RTB.

The additional complexity and cost of carrying a second sensor should better be used for flying two different  composition dipoles instead of   one. As argued in\,\cite{Adel2008,TBfocusIssue2012}, their measurements  can both be analyzed not only in the field of the Earth but also of the Sun and of dark matter at the center of our galaxy, thus avoiding  an accidental  cancellation of the charges of the test-body dipole or the attractor. This would increase the chance of finding a non-null result  and strengthen  its physical significance.

\section{Lessons from \ms\ and room for major  improvements}
\label{Sec:LessonsFromMicroscope}
With the zero-check sensor probably unable to firmly discriminate systematic errors, the final \ms\ test of the WEP will have to rely on standard procedures of systematic error checks in the different composition sensor, as envisaged before launch\,\cite{MicroscopeFocusIssue2012} when the plan was to  reach the mission  target $\eta_{_{\oplus}}=10^{-15}$  with an integration time corresponding to  $120$ orbits. In a $2$-yr mission duration there would be many such measurements  to check the result and  even improve it.  In the  words of \ms\ scientists\,\cite{MicroscopeFocusIssue2012}: ``\textit{The adopted trade-off remains on different sessions of 120 orbital periods. This is long enough to obtain the \eotvos\  parameter target exactitude of $10^{-15}$ in inertial mode and even better in rotating mode, by reducing the stochastic error with respect to the systematic evaluated one. This is also short enough to have time for many sessions with different experimental conditions.}'' 

It turns out that even in rotation mode, and at a faster rate than the maximum planned, the level  of noise over $120$ orbits is about 10 times  higher than expected, and  the entire mission duration is necessary  to reduce it  and  bring the precision of the WEP test closer to the original $10^{-15}$ target. Since systematic errors are not expected to disappear with more data, and all the integration time available is used to reduce random noise, their will be no time left to check them.

 We may therefore be left with the most sensitive test of the WEP ever, but no firm conclusion as to whether the equivalence principle  is violated or not.  Only another experiment in space, with higher precision,  shorter integration time and consequent reliable  systematic checks, could give the answer. The success of \ms, together with its limiting factors,  tell us that in orbit it is possible to reach a much higher precision, and give clear indications as to how to proceed in order to reach it.

High precision requires  low thermal noise, which results in a short integration time. This needs high $Q$ and high frequency of the signal to be obtained  by  rotating the spacecraft, the faster the better. As discussed in Sec.\,\ref{Sec:SuccessLimitations},  quality factors better than those of  the gold wires of \ms\  have been obtained, also at $\rm mHz$ frequencies. However,  this improvement alone would not be enough to reach a sensitivity to differential accelerations better than that  achieved by RTB. 

A solution often advocated for \ms\ is to eliminate any physical connection by replacing the gold wires  with an active system of electric discharging,  as in LPF. It would  increase  complexity and  cost, but  it is feasible. Once thermal noise from internal damping were eliminated by eliminating the suspensions, the next relevant one would be thermal noise from gas damping\,\cite{TrentoGasDamping2009,IntegrationTimePRD2014}. For the lowest possible residual pressure, a  way to reduce gas damping noise is by increasing the gap, up to $4\,\rm mm$ in the case of LPF, as compared with $600\,\mu\rm m$ in \ms.

Using the actual data for the test cylinders and the gaps, and assuming a residual pressure of $10^{-5}\,\rm Pa$  as planned before launch, we have calculated the differential acceleration noise due to gas damping in the SUEP sensor as it was done for GG\,\cite{IntegrationTimePRD2014}. The result is about $2\times10^{-13}\,\rm ms^{-2}/\sqrt{\rm Hz}$, while for GG  it was two orders of magnitude smaller, of about   $2.8\times10^{-15}\,\rm ms^{-2}/\sqrt{\rm Hz}$, thanks to the larger gaps ($2\,\rm cm$ vs $600\,\mu\rm m$) and to the higher mass of the test cylinders ($10\,\rm kg$ each vs $0.3$ and  $0.4\,\rm kg$ in SUEP). The value calculated for SUEP  is one order of magnitude smaller than  thermal noise from internal damping  in the gold wires estimated before launch\,\cite{MicroscopeFocusIssue2012},   and two orders of magnitude lower than the residual noise reported in orbit\,\cite{MicroscopePRL171204}.

The  capacitance is inversely proportional to the gap, so larger  gaps mean  a less  precise readout, which in fact in LPF is at nanometer level\,\cite{LPFcapacitors2017}, while \ms\ reports  $3\times10^{-11}\,\rm m/\sqrt{Hz}$ between $2\times10^{-4}\,\rm Hz$ and $1\,\rm Hz$\,\cite{MicroscopePRL171204}. For LPF this is not an issue because its main readout is based on laser interferometry\,\cite{LISAPF2016}. Instead, \ms\ relies  on the capacitance readout; moreover,  the capacitors   control the test cylinders, which would otherwise be unstable. Note that  gaps are  already  a factor of two larger than they were in GOCE, and a further increase is unlikely to yield a better net performance.   
 
Thus, eliminating the gold wires  is not going to  improve the \ms\  experiment; and the problem does not lie with the mechanical suspensions \textit{per se}, as the most precise and most successful gravitational experiments (RTB and gravitational wave detectors) demonstrate. Mechanical  suspensions at zero-g are perfectly predictable from ground measurements and there is nothing mysterious about using them in space (see \,\cite{Whirl1999}).
The stiffness can be predicted by semi-analytical, semi-numerical methods and verified experimentally in the required conditions of pressure and thermal stability and for the expected largest oscillation amplitude, by setting the flexure in oscillation in the horizontal plane so as not to be affected by local gravity. The quality factor at the frequency of interest is hard to predict, but it can be measured very reliably. There is certainly no need to fly a spring in order to establish its stiffness and quality factor!

Unlike electrostatic  suspensions, mechanical suspensions act as positive springs and  naturally  provide the restoring force needed by the test masses. For instance, the deflection of the torsion balance under the effect of a torque with a non-zero component along the suspension fiber, including that of a WEP violation, is counteracted by its torsional elastic constant.
 In \ms\ the restoring force must be provided actively, while the gold wire acts as an ancillary dummy spring with the sole purpose of ensuring electric grounding. Instead, mechanical suspensions can  provide both the restoring force and  electric grounding. 
The   torsion balance of the \eotwash\  WEP experiments weighs  only  $70$ grams in total, and  can be suspended with a very thin W wire of $20\,\mu\rm m$ diameter whose torsional elastic constant is very low (being inversely proportional to the  $4^{th}$ power of the thickness). In gravitational wave detectors the mirrors to be suspended are much heavier (about $40\,\rm kg$) and the fibers much thicker (about $350\,\mu\rm m$), but at  frequencies around $100\,\rm Hz$ thermal noise from internal damping is very low. Suspension fibers  are metallic in Virgo, soon to be replaced  with fibers in fused silica (as  in LIGO) with  even better quality factor.
In orbit  weight is no longer a limitation, and  large masses can be used, which reduces the effects of non-gravitational forces, including those due to thermal noise. Thus, in space tests of the WEP    mechanical suspensions are preferable. 

The solution we are led to is twofold. In the first place, we must use mechanical suspensions  with state-of-the-art fabrication, heat treatment and clamping procedures, so as to   ensure high $Q$.  Secondly, we must rotate the spacecraft much faster than \ms, so as to up-convert the signal to a much higher frequency, where thermal noise from internal damping  is significantly reduced.

 \ms\ has  demonstrated the advantage of space  for high precision rotating experiments. The possibility, unique to space, to spin the  entire ``laboratory'', that is the spacecraft,  along with the test cylinders makes rotation noise much lower in orbit than on ground.
However, the spin rate of \ms\  is limited by the need to rotate  around an axis which is not the symmetry axis of the test cylinders (\cite{MicroscopePRL171204}, Fig.\,1). So far the highest reported spin rate, achieved during SUEP 120-orbit run which has given the result (\ref{Eq:etaPtTi}), has been of $2.94\times10^{-3}\,\rm Hz$.  
Spacecraft can spin much faster than that, and be passively stabilized by rotation around the symmetry axis. 

Mechanical suspensions are very versatile and allow the concentric test cylinders to  be arranged  in such a way that they  co-rotate with  the spacecraft around the symmetry axis, being sensitive in the plane perpendicular to it. In this  plane the relative displacements caused by tiny low frequency differential accelerations between the test cylinders --such as a violation signal-- can be detected, up-converted  by rotation to a much higher  frequency where thermal noise  is much lower\,\cite{PRLthermalnoise}.
After  initial spin up, spacecraft stabilitazion  is maintained passively by conservation of angular momentum, which ensures extremely low rotation noise and  does not need  propellant --to be left for drag-free control and occasional manoeuvres\,\cite{GGssoPLA2003}.
At $\rm Hz$ rather than $\rm mHz$ frequency thermal noise from internal damping is very low, and gaps can be increased so as to reduce also gas damping noise and make the integration time short even  for a very  high precision target\,\cite{IntegrationTimePRD2014}.
With  very low thermal noise a readout of comparable low noise is needed. With cm level gaps a capacitance readout is not sensitive enough; a laser gauge with very low noise at $1\,\rm Hz$  is well feasible\,\cite{LIG2015,LIG2016}.

\begin{figure}[H]
\begin{center}
\includegraphics[width=0.25\textwidth]{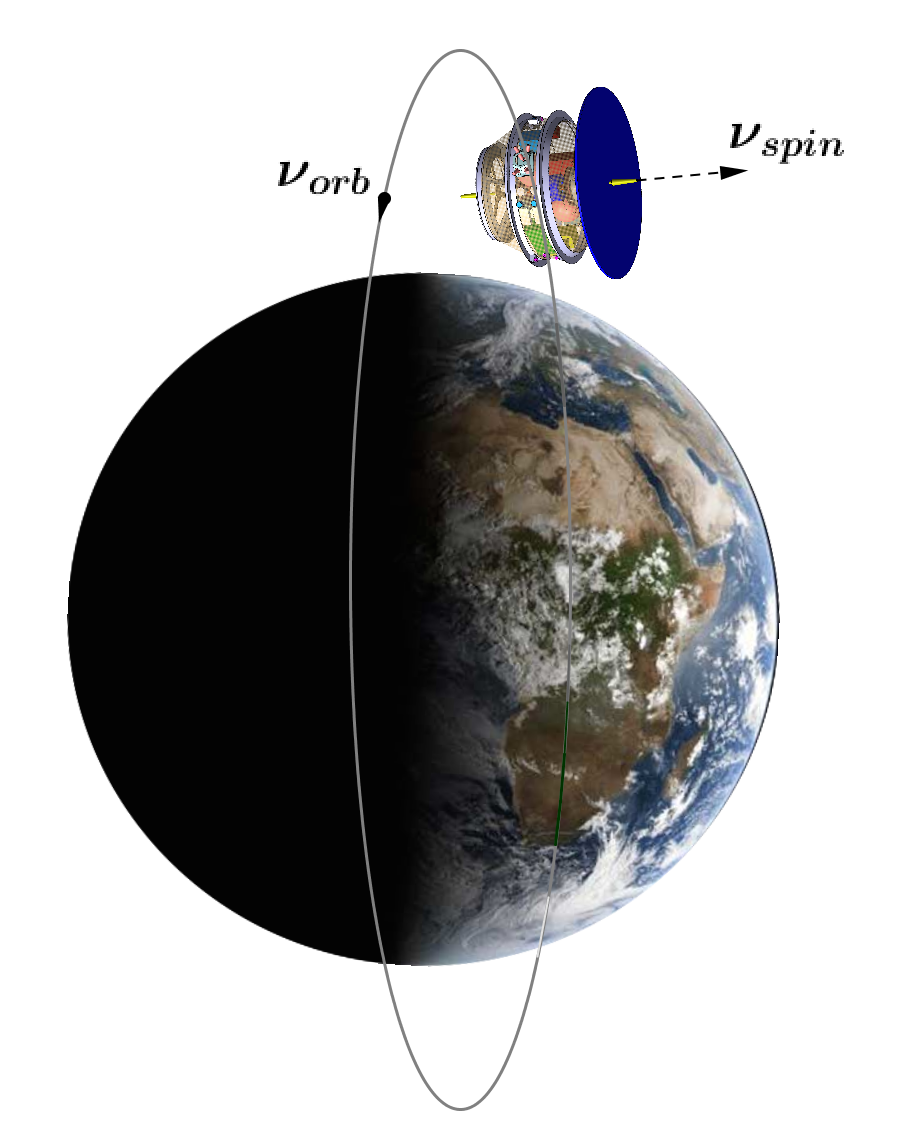}
\caption{Sketch of the GG satellite, with cylindrical symmetry and the dish of solar cells facing the Sun, in a high inclination sun-synchronous orbit ($\nu_{orb}\simeq1.7\times10^{-4}\,\rm Hz$) similar to that of \ms.   The spacecraft is passively stabilized by one-axis rotation around the symmetry axis at $\nu_{spin}=1\,\rm Hz$. After initial spin-up relative to inertial space rotation is maintained by conservation of angular momentum. (Figure not to scale; the bulk of the spacecraft body has a diameter of about $1.5\,\rm m$).}
\label{Fig:GGinOrbit}
\end{center}
\end{figure}
\vskip-6mm

The proposed GG space experiment incorporates all the features suggested by the \ms\ experience  and aims  to test the  WEP to $1$ part in $10^{17}$ (\cite{PRLthermalnoise,GGfocusIssue2012,GGwebpage,GGssoPLA2003}, \cite{BraginskyPaper2017}).
%

Like \ms, it will fly in a low altitude, high inclination, sun-synchronous orbit (with orbital frequency $\nu_{orb}\simeq1.7\times10^{-4}\,\rm Hz$). However, thanks to its cylindrical symmetry (built around the concentric test cylinders), it can be  passively stabilized by one-axis rotation around the symmetry axis at the rotation rate  $\nu_{spin}=1\,\rm Hz$ (Fig.\,\ref{Fig:GGinOrbit}). 
  The coaxial, concentric test cylinders  are located at the center of mass of the spacecraft, co-rotate with it  and are sensitive to differential forces acting  in the plane perpendicular to the spin/symmetry axis. The relative displacements caused by any such force, like a violation signal, are read by a laser interferometry gauge, also co-rotating with the whole system. In the non-rotating frame of the spacecraft the  violation signal is at the orbital frequency; the laser gauge rotating at $\nu_{spin}\gg\nu_{orb}$ reads it at  $\nu_{spin}$, where thermal noise from internal damping is much lower than it would be at $\nu_{orb}$, making the integration time much shorter\,\cite{PRLthermalnoise,IntegrationTimePRD2014}. This is the key to reaching a very high precision. It suffices to  notice that improving by  a factor of $10$ in sensitivity requires --with a given level of thermal noise-- an integration time $100$ times longer. As an example, \ms\ early test (\ref{Eq:etaPtTi})  has required $8.26$ days of integration time; were it aiming at  $10^{-17}$, the same experiment would need about $6.7$ million days for one single measurement.
  
The expectations for GG, based on theoretical analysis, numerical simulations and laboratory tests, are for a signal-to-noise ratio of 2 in a few hours\,\cite{IntegrationTimePRD2014}. This allows a WEP test to $10^{-17}$ to be completed in $1\,\rm d$ (about $15$ orbits). Then, since the spin axis (and the sensitive plane perpendicular to it)  are fixed in inertial space, while the nodal line of the sun-synchronous orbit moves  by about $1^{\circ}/\rm d$ (for the solar panel to follow the annual motion of the Sun), a large number of $1$-d runs shall be available, in different physical conditions provided by the dynamical evolution, to allow a violation signal at $10^{-17}$ level to be separated with certainty from systematic errors\,\cite{Moriond2011NullChecks,NobiliMoriond2011}.

As discussed in Sec.\,\ref{Sec:SuccessLimitations}, for a test of the WEP in orbit to reach a very high precision it must deal with the huge effect of drag which requires, in addition to partial compensation by drag-free control, also a high level of Common Mode Rejection (CMR).  The largest common mode effect in orbit is the inertial acceleration equal and opposite to the acceleration of the spacecraft caused by non-gravitational forces such as  air drag and solar radiation pressure. A high level of CMR can be achieved if the test masses are arranged as a balance. This is how torsion balances have defeated mass dropping tests by many orders of magnitude. A torsion balance as such is not suitable for space\,\cite{BraginskyPaper2017}. However, space is  favorable for the realization of  a  very sensitive balance because in orbit the largest common mode force against which the balance is balanced  is many orders of magnitude weaker than $1$-$g$ on ground (a space version of the Watt balance has been considered at PTB, the German national metrology institute\,\cite{DittusWattBalance}).

In order to achieve a high level of CMR the coaxial test cylinders in GG are arranged to form a beam balance, with the beam along the spin/symmetry axis, hence sensitive to differential forces in the plane perpendicular to it. The peculiarity of the GG  balance is that, unlike ordinary beam balances, the two masses are concentric, which is a crucial requirement in space tests of the WEP   to minimize classical differential tidal effects. The way such a beam balance with concentric test masses can be realized  is based on an ingenious combination (originally designed by D. Bramanti) of weak, high $Q$ flexures made in CuBe and coupling arms (with special attention to symmetry considerations) whose lengths can be finely adjusted to reach a very good balancing against the common mode inertial acceleration caused by drag.  An animation of this balance is available on the front page of the GG website~\cite{GGwebpage}. A $1$-$g$ version of it has been realized and tested in the lab with the ``GG on Ground'' (GGG) demonstrator\,\cite{GGfocusIssue2012},\,\cite{BraginskyPaper2017}.
A second concentric sensor can be accommodated inside the GG spacecraft, whose test cylinders are arranged as a similar balance and  can take data simultaneously with the first one\,\cite{GG2003PLA}. As argued in Sec.\,\ref{Sec:SUEPvsSUREF}, it should better be a second composition dipole than an equal composition sensor.

GG will check \ms's final result with at least $2$ orders of magnitude better precision and improve by $4$ orders of magnitude over RTB tests, thanks to the stronger  signal in orbit (by about  $500$ times) and to a lesser extent by fully exploiting the advantages  of space in order  to reach a sensitivity to differential accelerations  better than RTB (by $20$ times). The latter factor would be an improvement also of  RTB tests of the WEP relative to the Sun and to dark matter in our galaxy.

\section{Conclusions}
\label{Sec:Conclusions}
MICROSCOPE demonstrates that, by going to space, a WEP experiment is within reach that can truly address the foundations of physics. For this potential to become real, numerous lessons from MICROSCOPE must be learned. 

In this paper we have shown that MICROSCOPE is limited by thermal noise due to the low quality factor of the gold wires used for electrical grounding, and yet outstanding results were achieved by rotating the spacecraft faster than planned. This fact establishes spacecraft rotation as the most effective way of improving the precision of WEP tests. Further, we have shown that the Pt-Pt ``zero-check'' sensor is less sensitive to a wide class of systematic effects than the Pt-Ti sensor, and therefore can neither confirm nor disprove a violation signal at the achieved sensitivity level. Thus, there is no alternative to a rigorous campaign of systematic checks made up of many measurements in different experimental conditions, requiring short integration time, hence low thermal noise. 

All these facts point the way to the design required of a breakthrough experiment. A new experiment must feature high quality, state-of-the-art mechanical suspensions as demonstrated in torsion balance ground tests of the WEP  and in gravitational wave detectors. For low thermal noise it must up-convert the signal to much higher frequency, which is easily achieved in a fast rotating spacecraft, provided its cylindrical symmetry replicates that of the test bodies in a Russian doll setup with rotation around the symmetry axis. Against the huge effect of drag it must provide  a  high level of common mode rejection by exploiting in absence of weight the versatile nature of mechanical suspensions --whose properties can be predicted and measured in the lab-- so that they become the solution rather than being the problem. It must be free by design from systematic effects due to differential gas pressure (the radiometer effect), and free from the ubiquitous constraint of narrow gaps (to reduce gas  damping noise and electric patch effects) by replacing the capacitance readout with laser  interferometry. 
To check systematics, it must allow multiple measurements, as are made possible by low thermal noise and short integration times. 

For all of these aspects, viable solutions exist based on proven technology.  The proposed GG experiment aims to test the WEP to $10^{-17}$ at room temperature and incorporates all the required features (\cite{PRLthermalnoise,GGfocusIssue2012,GGwebpage,GGssoPLA2003}, \cite{BraginskyPaper2017}).  All future proposals for WEP experiments in space will be confronted with the issues raised by \ms, and will be compared with the solutions offered by GG.

 \textbf{Acknowledgements.}  Thanks are due  to the colleagues of the GG collaboration for their contributions. The  input for this work came during the 656th Wilhelm und Else Heraeus Seminar on ``Fundamental Physics in Space'' held in Bremen  on October 2017. A.M.N wishes to thank the organizers for their invitation. Special thanks to Manuel Rodrigues, of the \ms\ collaboration, for useful information and discussions.

\end{document}